\documentclass[aps,prl,twocolumn,showpacs,superscriptaddress,groupedaddress]{revtex4-1}  
\usepackage{graphicx}  
\usepackage[font=small,labelfont=bf]{caption} 
\usepackage{dcolumn}   
\usepackage{color}	
\usepackage{bm}        
\usepackage{amssymb}   
\usepackage{mathrsfs}

\newcommand{\todo}[1]{}
\renewcommand{\todo}[1]{{\color{red} TODO: {#1}}}

\begin{document}

\widetext

\title{Efficient Simulation of Tunable Lipid Assemblies Across Scales and Resolutions}
\author{John M. A.~Grime}
\affiliation{Emerging Technologies Team, University of Oklahoma, Norman, Oklahoma, United States of America}

\author{Jesper J.~Madsen}
\email{jespermadsen@usf.edu}
\thanks{Corresponding author}
\affiliation{Department of Global Health, College of Public Health, University of South Florida, Tampa, Florida, United States of America}
\vskip 0.25cm
\date{\today}

\begin{abstract}
We present a minimal model for simulating dynamics of assorted lipid assemblies in a computationally efficient manner. Our model is particle-based and consists of coarse-grained beads put together on a modular platform to give generic molecular lipids with tunable properties. The interaction between coarse-grained beads is governed by soft, short-ranged potentials that account for the renormalized hydrophobic effect in implicit solvent. The model faithfully forms micelles, tubular micelles, or bilayers (periodically infinite, bicelles, or vesicles) depending on the packing ratio of the lipid molecules. Importantly, for the self-assembled bilayer membranes it is straightforward to realize gel and fluid phases, the latter of which is most often of primary interest. We show that the emerging physics over a wide range of scales and resolutions is (with some restrictions) universal. The model, when compared to other popular lipid models, demonstrates improvements in the form of increased numerical stability and boosted dynamics. Possible strategies for customizing the models (e.g., adding chemical specificity) are briefly discussed. An implementation is available for the LAMMPS molecular dynamics simulator [Plimpton. J. Comp. Phys. \textbf{117}, 1-19. (1995)] including illustrative input examples from the simulations we present. 
\end{abstract}

\maketitle


\section{\label{sec:intro}I. INTRODUCTION}
Above a certain critical concentration, amphiphilic lipid molecules in water assemble into large-scale structures and arrangements, including membranes, that allow the hydrophobic "tails" to be shielded from the water by creating an interface where the hydrophilic "heads" are displayed toward the water phase. Plasma membranes in particular are essential in biology where they facilitate compartmentalization of distinct chemical environments between the fundamental unit of life and its surroundings as well as within the cell itself. 

Because of the ubiquity of lipids membranes in biology, it has long been desirable to use theoretical and computational methods to explore their behavior and calculate properties efficiently. While there are many approaches toward this aim (see, e.g., ref. \cite{radha} and the references within), major conceptual progress was made more than a decade ago when it was discovered that solvent-free lipid models with simple pairwise potentials could give rise self-assembled fluid bilayer membranes with tunable properties \cite{farago,bpb,ckd}. Arguably, the  $1/r^{12}$ hard-core repulsive term used in the potential energy functions of early models is adopted mainly for historical reasons and not appropriate at the coarse-grained resolution(s) typically employed in these models. Careful considerations affirm that soft and squishy interaction potentials are correct at these resolutions \cite{dpd1,dpd2,dpd3,dpd4}. One straightforward way to overcome this is to use a smaller power on the repulsive part of the potential, as done in e.g. the Shinoda-DeVane-Klein (SDK) model \cite{sdk}. However, at highly coarse-grained representations, the Lennard-Jones functional form still has undesirable properties that makes its use unfeasible. There are several ways in which such shortcomings can be remedied. For instance, one can formulate approximations for the thermodynamic equations of state and use density expansions of the excess free-energy functional directly \cite{mueller}. Another recent solution was implemented in the model by Revalee \textit{et al.} \cite{sunilkumar}, which uses a functional form with finite pairwise repulsion at small distances, $r$: The repulsive force increases linearly as $r \rightarrow 0^+$ (below a characteristic size where the repulsive and attractive regime is stiched together) in a similar fashion to the soft conservative pariwise repulsive force used in Dissipative Particle Dynamics \cite{dpd,smit,guo}. This effectively tries to capture "pressure" between different particles (in a qualitative sense, at least). The model we here propose is conceptually similar in that we use short-ranged, soft-core potentials. On this basis, we construct a family of generic lipid models with tunable properties for studying membrane-associated biological processes with their remarkable complexity from a minimalistic and physics-based standpoint.

The article of organized as follows: We begin by describing the topology of the lipid models we here investigate and the potentials that govern their interactions. Self-assembly properties are elucidated for all models and we show examples of large-scale morphological transitions. For the 4-bead model, we elucidate the properties of the formed fluid bilayer membranes in some detail. Finally, for the most efficient of all presented models, the quasi-monolayer model, we show how properties of the fluid bilayer membranes depend on the intrinsic length scale and demonstrate their flexibility by performing \textit{in silico} pulling experiments on a vesicle. 

\section{\label{sec:models} II. MODEL AND METHODS}
In the present work, we describe lipid molecules as semiflexible linear chains of 2-5 beads per lipid. We furthermore describe a quasi-monolayer model with "1.5" beads/lipid (i.e., 3 beads per bilayer segment) (Fig. \ref{fig:models}). The hydrophobic effect acting on the amphiphilic lipid is heuristically renormalized into interfacial and/or tail cohesion, depending on the lipid resolution. The rigidity of the lipid molecule is maintained by appropriate 2-body (bond) and 3-body (angle) intramolecular potentials. 

\begin{center}
\includegraphics[width=0.35\textwidth, angle=0]{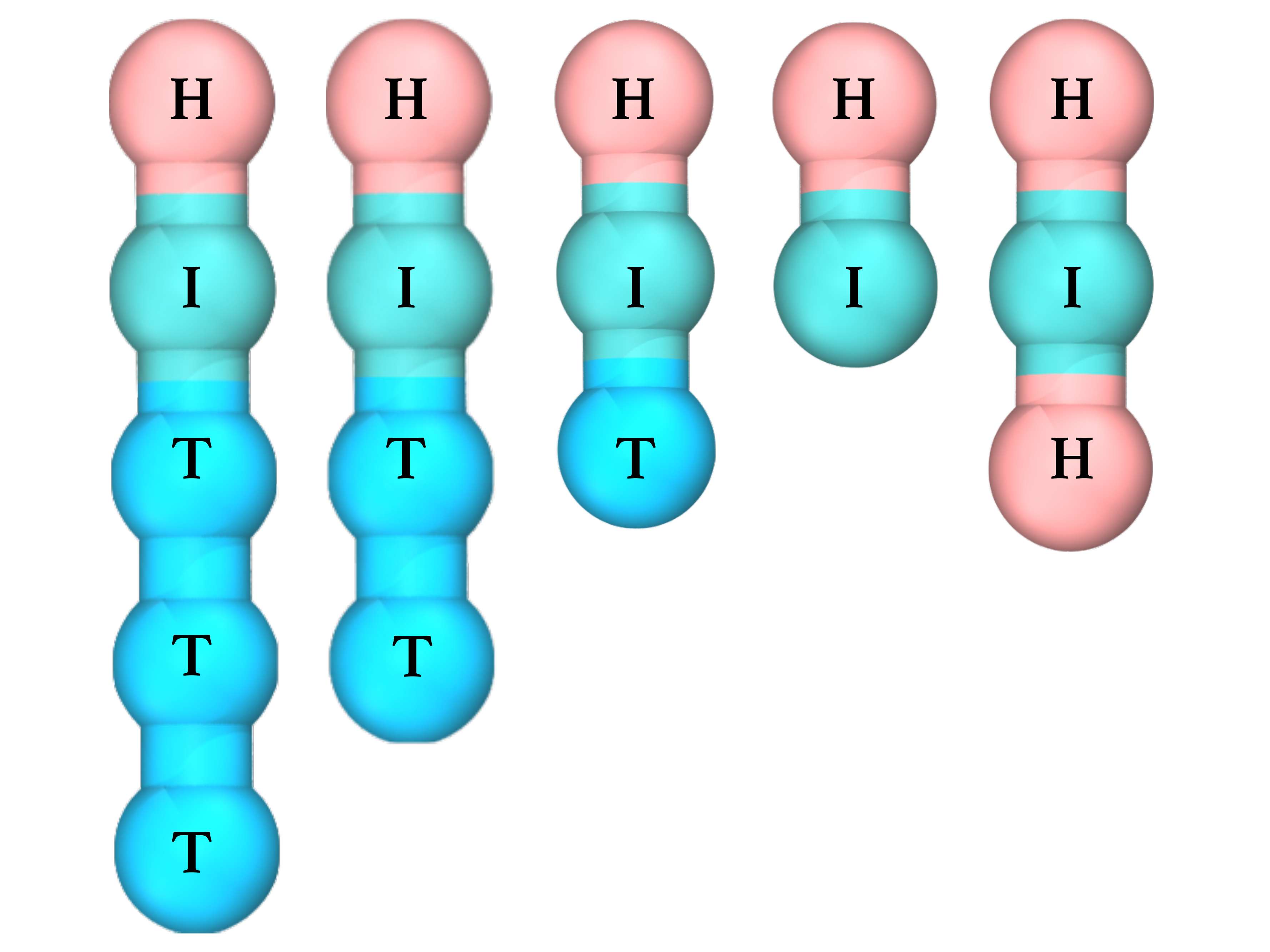}
\captionof{figure}{The lipid models described here. From left to right: 5-bead model, 4-bead model, 3-bead model, 2-bead model, and the quasi-monolayer ("1.5 bead") model}
\label{fig:models}
\end{center}

The total potential energy function is defined as

\begin{eqnarray} U(\{\textbf{r}_i\})&=& \sum_{i>j} U_{pair}(r_{ij}) \nonumber \\
&&+ \sum_{i}U_{bond}(r_{i,i+1}) \nonumber \\
&&+ \sum_{i}U_{angle}(\theta_{i,i+1,i+2}), \end{eqnarray}
where $r_{ij}=| \textbf{r}_i-\textbf{r}_j |$ and summation indices run over all unique bonds ($i,i+1$), angles ($i,i+1,i+2$), and non-bonded pairs ($i>j$) of the system, respectively. Bond and angle potentials are both described using simple harmonic functions,

\begin{eqnarray}
U_{bond}(r)&=&K_b(r-r_0)^2 \\
U_{angle}(\theta)&=&K_\theta (\theta-\theta_0)^2,
\end{eqnarray}
while the non-bonded interactions are governed by the following soft pair potential (Fig. \ref{fig:potential})

\begin{equation}
\begin{array}{cc}
U_{pair}(r)=\\ 
		\left \{ 
                \begin{array}{ll}
                  -\left( \frac{A}{a}  \right) \sin (r \cdot a),& \text{if }   r \leq r_0\\
                  -\left( \frac{B}{b}  \right) \sin \left( \frac{\pi}{2} + (r-r_c) \cdot b \right),&\text{if }  r_0 < r \leq r_c \\
                  0,& \text{otherwise,}\\
                \end{array} \right. \\
\end{array}
\end{equation}
with the constants $a=\frac{\pi}{2r_0}$ and $b=\frac{\pi}{r_c-r_0}$. Therefore, the forces, $F_{pair}(r)= -\nabla U_{pair}(r)$, are conveniently expressed as

\begin{equation}
F_{pair}(r)=  \left \{ 
                \begin{array}{ll}
                  A \cos \left(  r\cdot a \right) ,& \text{if }  r \leq r_0\\
                  B \cos \left(  \frac{\pi}{2} + (r-r_c) \cdot b  \right) , & \text{if }  r_0 < r \leq r_c  \\
                  0, & \text{otherwise.}\\
                \end{array}
\right.
\end{equation}

\begin{center}
\includegraphics[width=0.235\textwidth, angle=0]{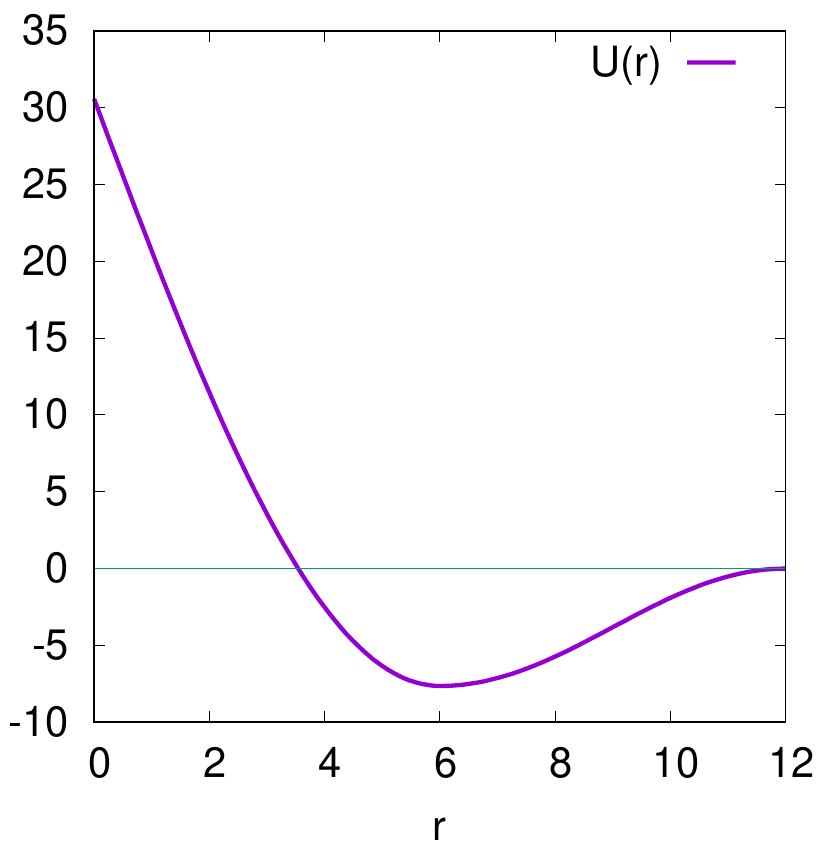}
\includegraphics[width=0.235\textwidth, angle=0]{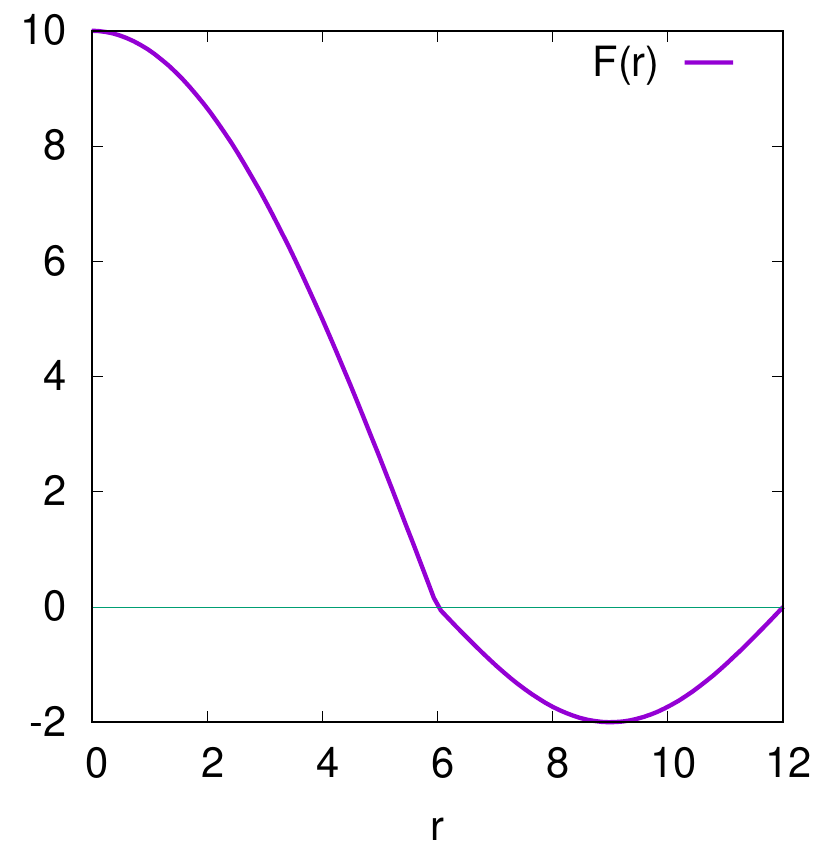}
\captionof{figure}{Pairwise interaction potential (\textit{left}) and force (\textit{right}) plotted for the values $A=10$, $B=2$, $r_0=6$, and $r_c=12$.}
\label{fig:potential}
\end{center}

The pair potential is piecewise smooth and at least $\mathscr{C}^1$-continuous on the open interval $(0,r_c)$ for all non-trivial parameter choices ($\{ A,B,r_0,r_c\in \mathbb{R}_{>0}^{4}\ |\ r_c\textgreater r_0\textgreater 0 \}$), which, in our experience, is sufficient to give rise to well-behaved dynamics in all tested situations where intuitively reasonable parameters are used \footnote{Parameters can obviously be chosen in order to make the potential $\mathscr{C}^2$-continuous. While this is an unnecessarily strict restriction for practical applications, we do advice that care is taken if the discontinuity in the derivative of the force is "large." One hypothetical situation where a force-derivative discontinuity is useful is if a weak, long-ranged attractive basin is needed in combination with very low compressibility above a certain threshold pressure. However, in such situations there may be better alternatives.}. 

Molecular dynamics (MD) is simulated using the Langevin formulation,

\begin{eqnarray}
M\ddot{r}&=&-\nabla U(r)-\Gamma\dot{r}+R(t),\\
\langle R(t)\rangle &=&0,\\
\langle R(t).R(t')\rangle &=& 6k_BT \Gamma \delta_{i,j} \delta (t-t'),
\end{eqnarray}
where $-\nabla U(r)$ is the force field, $\dot{r}$ is the velocity, $\ddot{r}$ is the acceleration, $T$ is the temperature, $\Gamma$ is the friction coefficient, $k_B$ is the Boltzmann constant, and $R(t)$ is the random force. The thermostat coupling is chosen to be very weak ($T=300 K$; $T_{damp}=100,000 \text{fs}$) to reduce the influence of the thermostat on the natural dynamics of the system. For constant-pressure simulations, we use the Nos\'{e}-Hoover barostat with Martyna-Tobias-Klein correction \cite{mtk} to realize the tensionless ensemble ($P_{ext}=0$; $P_{damp}=100,000 \text{fs}$); coupling between the $x-$ and $y-$dimensions were used for simulations of periodically infinite bilayers. We remark that hydrodynamic interactions, including simple volume effects, are not correctly captured unless special steps are undertaken to incorporate the role of the solvent \cite{lenz,atzberger}, which we shall not pursue here. Unless explicitly stated otherwise, we use the following parameters: $\tau = 50 \text{fs} $ timestep for integration, $K_b=25k_BT${\textit \AA}${}^{-2}$, $A=25k_BT${\textit \AA}${}^{-1}$, head/interface/tail bead sizes $r_0=\{0.75R,R,R\}$, pair cutoff $r_c=2R$, and $R=7.5${\textit \AA}. A standard excluded-volume repulsion is employed between all pairs ($U_{pair}$ with $B=0$). The interfacial/tail cohesion can be distributed equally over all sticky (non-head group) beads \textit{via} the $B$ parameter according to Table \ref{tab:parameters}. For simplicity, the cohesion only acts within each bead type (interface or tail) with all cross-interactions being zero, $B_{cross}=0$. All simulations are performed using the LAMMPS MD simulator \cite{lammps} and example input parater files for fluid bilayers of all presented models can be found in the supplemental material online \cite{supplstuff}.

\begin{center}
\begin{tabular}{ |p{1.5cm}||p{1.1cm}|p{1.1cm}|p{1.1cm}|p{1.1cm}|p{1.6cm}|  }
 \hline
 & \multicolumn{5}{|c|}{Lipid model} \\
 \hline
 Cohesion, $B$ param. & 5-bead & 4-bead & 3-bead & 2-bead & Quasi-monolayer \\
 \hline
Intf-intf & $\frac{1}{R}\frac{7.0}{4}$ & $\frac{1}{R}\frac{7.0}{3}$ & $\frac{1}{R}\frac{7.0}{2}$ & $\frac{1}{R}\frac{5.0}{1}$ & $\frac{1}{R}\frac{7.5}{1}$ \\
\hline
Tail-tail & $\frac{1}{R}\frac{7.0}{4}$ & $\frac{1}{R}\frac{7.0}{3}$ & $\frac{1}{R}\frac{7.0}{2}$ & N/A & N/A \\
 \hline
otherwise & $0$ & $0$ & $0$ & $0$ & $0$ \\
 \hline
\end{tabular}
\captionof{table}{Example parameters for fluid bilayers. Reported values are in units of $k_BT$\textit{\AA}${}^{-1}$. ``Intf'' = interface. ``N/A'' = Not Applicable. }
\label{tab:parameters}
\end{center}

Fluid membranes in the tensionless ensemble ($\gamma=0$) will exhibit height fluctuations (i.e., undulations) according to the equation

\begin{eqnarray}
S_u(|\vec{q}|)&\equiv& N \langle |u(\vec{p})|^2 \rangle \\
&=& \frac{k_BT}{ \langle a \rangle (\gamma |\vec{p}|^2 + k_C |\vec{p}|^4 + \mathcal{O}(|\vec{p}|^6)) } \\
&\approx&\frac{k_BT}{\langle a \rangle k_C |\vec{q}|^4}, \label{eq:spectrum}
\end{eqnarray}
which can be derived from the Canham-Helfrich Hamiltonian \cite{canham,helfrich} using Fourier analysis and the equipartition theorem in the Monge gauge. Hence, eq. \ref{eq:spectrum} provides a convenient method for estimating the bending modulus from equilibrium simulations of tensionless bilayers via the undulation spectrum \cite{spectra}. 

\section{\label{sec:results} III. RESULTS AND DISCUSSION}

\noindent {\bf Self-assembly of lipid nano-structures and their morphology}

\begin{center}
\begin{figure*}
\includegraphics[width=0.99\textwidth, angle=0]{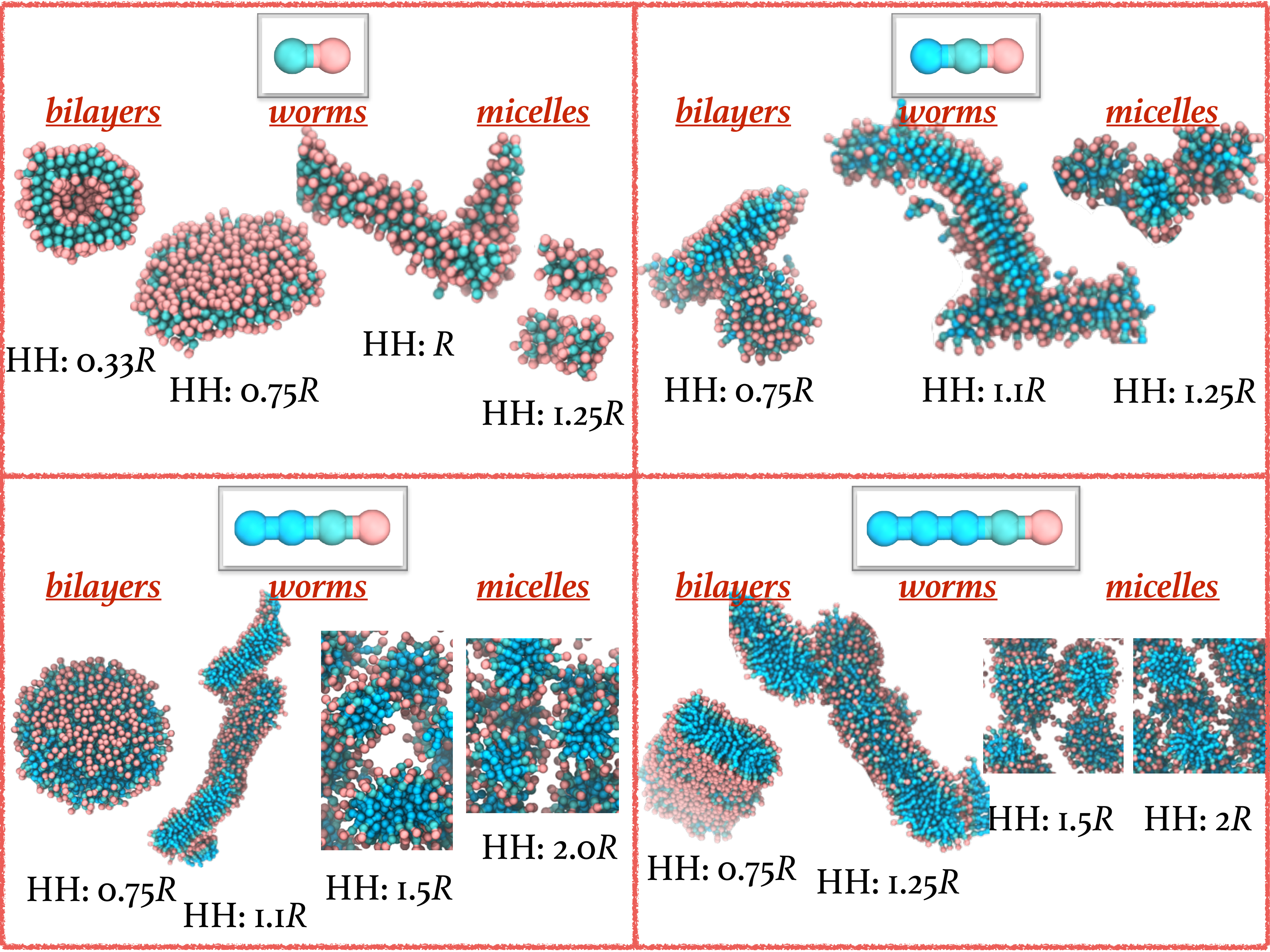}
\captionof{figure}{Self-assembled nano-structures reveal a remarkably conserved space of accessible morphologies from bilayers to worms and micelles. 450 lipids were simulated for $0.5\times 10^6$ MD steps. A cutting plane is used in some of the snapshots to expose the internal structure of the assemblies. Movies corresponding to the snapshots are available in the supplemental material online \cite{supplstuff}. }
\label{fig:assembly}
\end{figure*}
\end{center}

The packing ratio, or how conical the molecular lipid is, is the major determinant of the emerging morphology of self-assembled lipid nano-structures \cite{israel}. We find that a generic series of morphologies form upon tuning the relative size of the (purely repulsive) head bead essentially independent of how the internal degrees of freefrom of the lipid molecules are resolved (Fig. \ref{fig:assembly}).


If a vesicle is constructed with a diameter that is too small, a morphological transition into a planar bilayer can be observed to spontaneously occur (Fig. \ref{fig:unfurl}). This happens because it is energetically favorable to flatten the membrane and this is sufficient to overcome the line tension associated with a free-standing membrane edge \cite{nielsen}.

\begin{center}
\includegraphics[width=0.48\textwidth, angle=0]{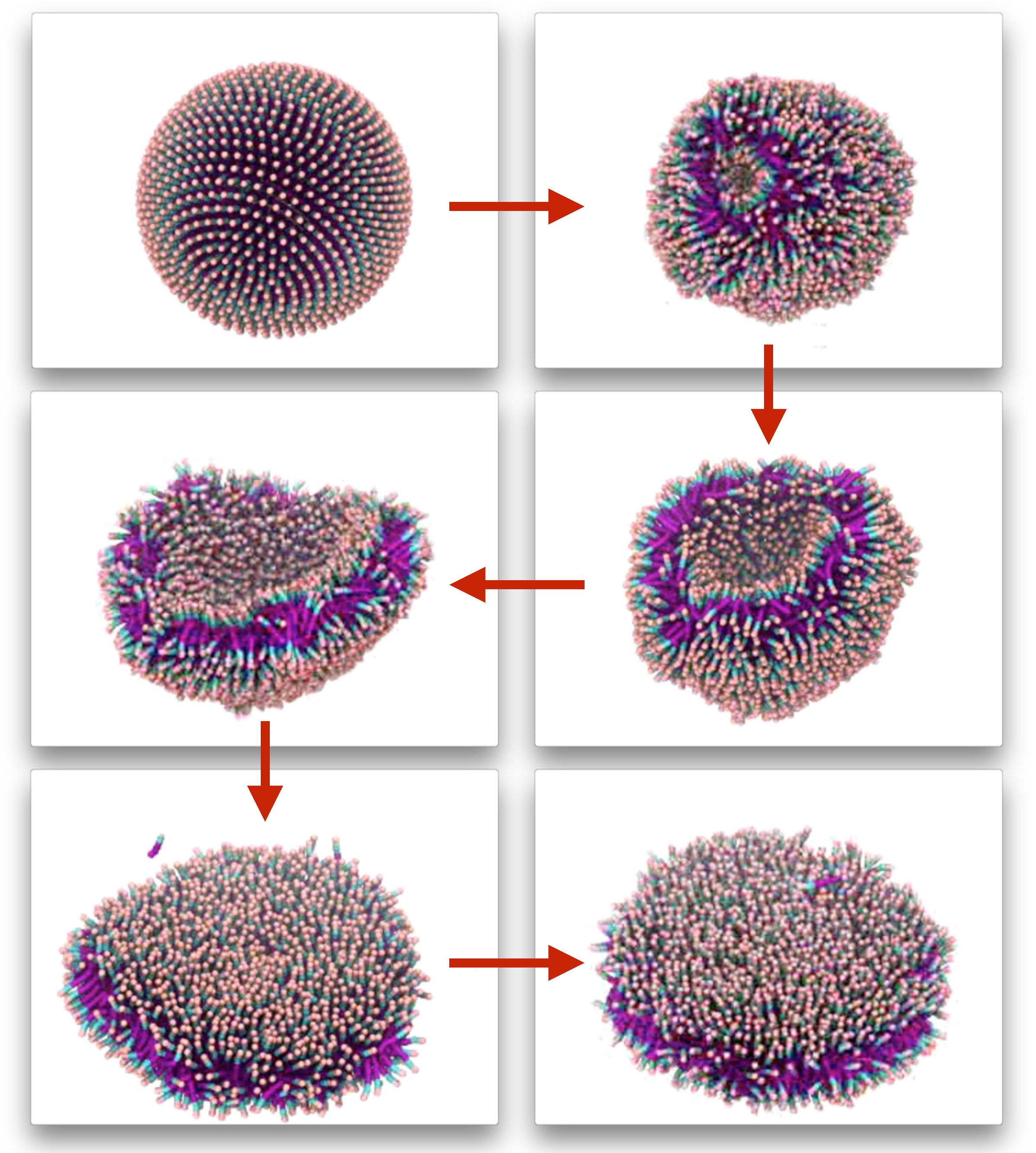}
\captionof{figure}{Simulation snapshots of the "mouth-opening" morphological transition that transforms the initialized vesicle of 5-bead lipids into a planar disc bicelle. A movie corresponding to the snapshots is available in the supplemental material online \cite{supplstuff}.}
\label{fig:unfurl}
\end{center}

\noindent {\bf Properties of the 4-bead model}

As an example, we shall characterize the properties of the 4-bead model and remark that similar trends are, again, seen for all discussed models. The projected area per lipid (APL) of the bilayer membrane of 4-bead lipids can be seen in Fig. \ref{fig:4site-apl}. 

\begin{center}
\includegraphics[width=0.48\textwidth, angle=0]{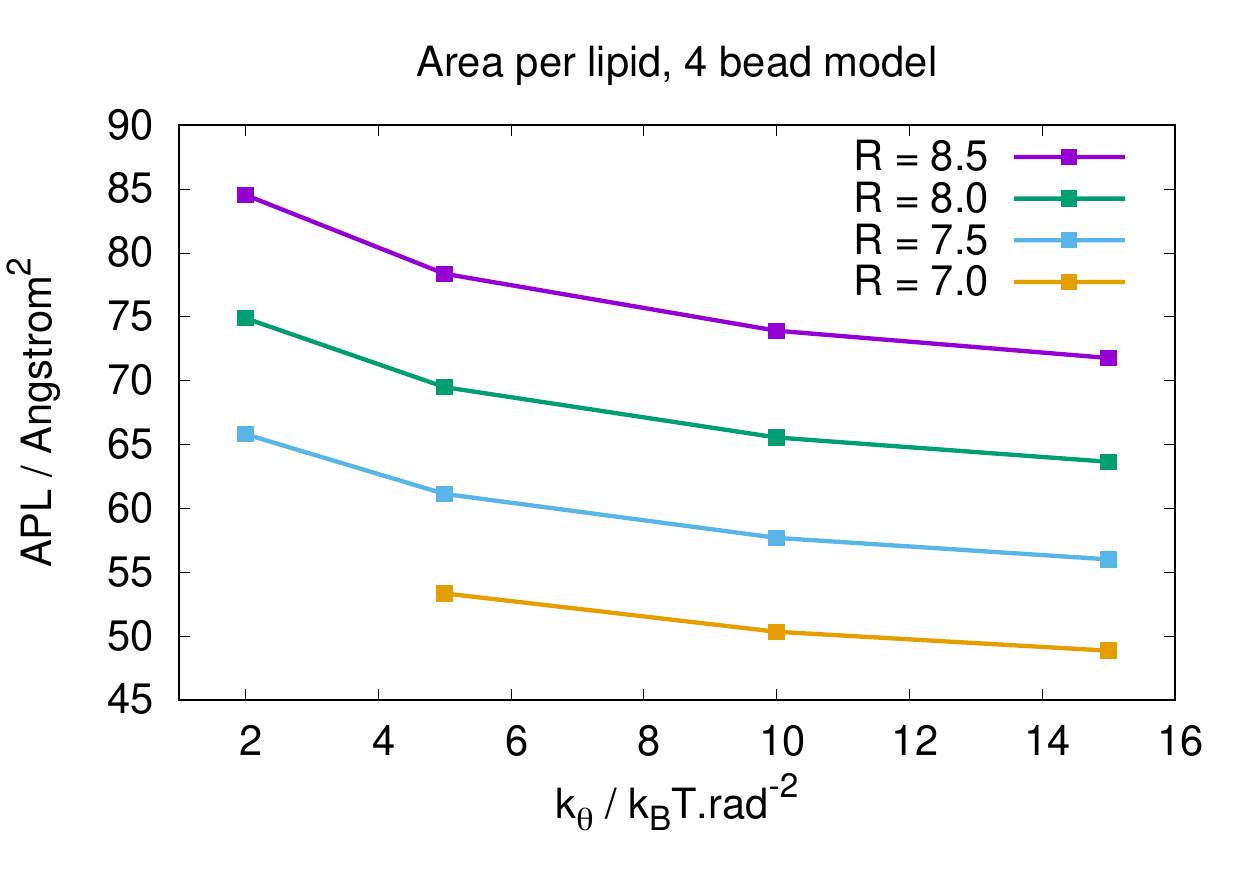}
\captionof{figure}{Projected area per lipid (APL) plotted for the 4-bead model as function of the angle potential parameter, $K_\theta$, for bead sizes $R=\{7.0, 7.5, 8.0, 8.5\}$. 20,000 lipid molecules were simulated for $50\times 10^6$ MD steps. Standard errors (of the means) are smaller than the plotted symbols.}
\label{fig:4site-apl}
\end{center}

The corredponding bending modulus, $k_C$, for the membranes in these simulations is plotted in Fig. \ref{fig:4site-bend}. The bending modulus increases monotonically with the angle potential parameter of the lipid model, which is expected since stiffer lipid molecules should give rise to a stiffer membrane. There is little to no dependence of the bending modulus on the intrinsic size scale, $R$, over the tested range. 

\begin{center}
\includegraphics[width=0.48\textwidth, angle=0]{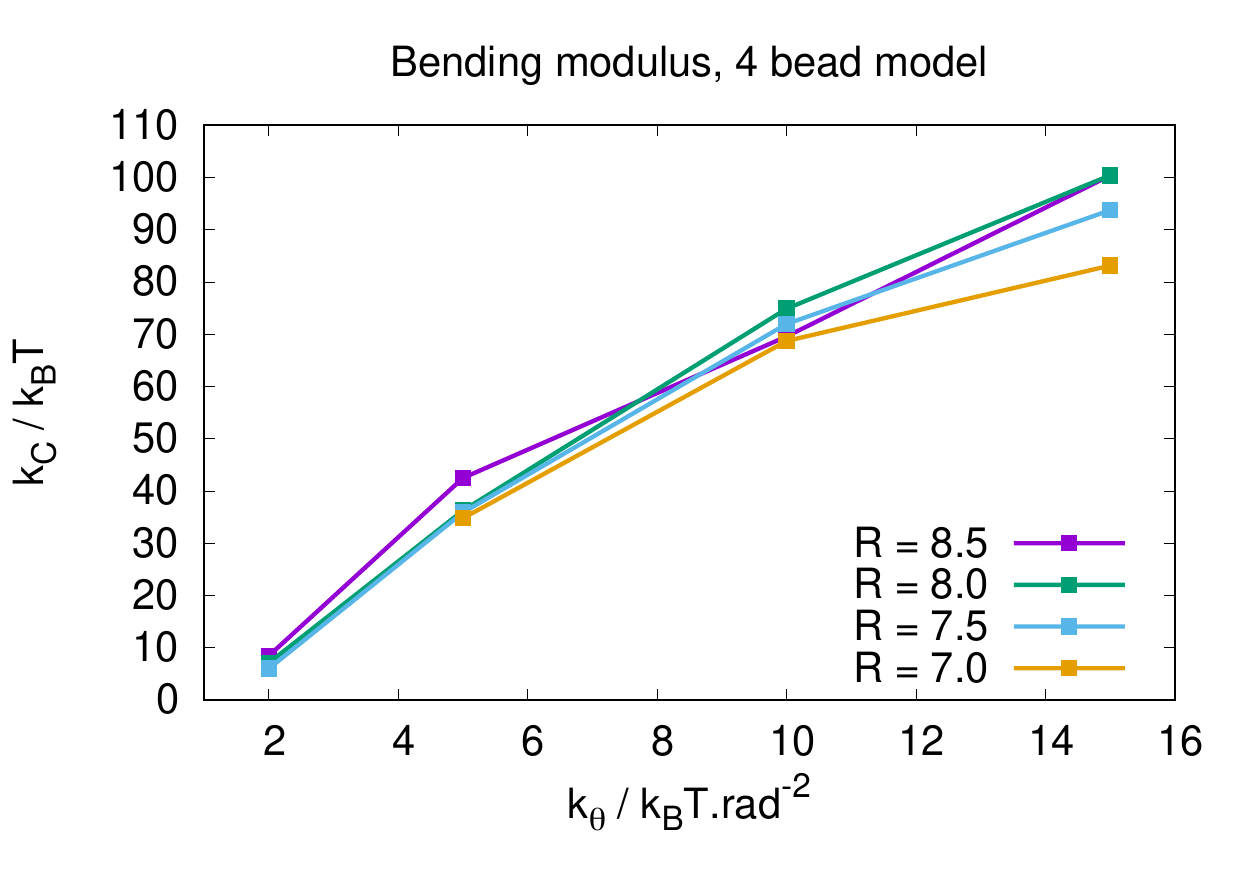}
\captionof{figure}{Bending modulus, $k_C$, plotted for the 4-bead lipid model as function of the angle potential parameter, $K_\theta$, for bead sizes $R=\{7.0, 7.5, 8.0, 8.5\}$. 20,000 lipid molecules were simulated for $50\times 10^6$ MD steps. Statistical errors in the least squares fits are smaller than the plotted symbols.}
\label{fig:4site-bend}
\end{center}

Recall, however, that invoking periodic boundary conditions, as we here do, in-itself affects the bending modulus of the membrane by effectively pinning it at the box boundary and truncating undulation modes whose wavelength exceeds the simulation box length, $L$. This gives rise to a system-size dependent bending modulus (Fig. \ref{fig:4site-bend-size}). The membrane is seen to get stiffer as the angle potential parameter, $K_\theta$, increases and the lipids get more rigid. Furthermore, for a chosen value of $K_\theta$, the membrane is supposed to get softer as the system size increases. The reason that these trends appear to be violated in Fig. \ref{fig:4site-bend-size} is most likely that bending modulus estimates from height fluctuation spectra are less reliable for small system sizes where methods based on deformation response might do a better job \cite{bend-rigatoni,bend-deserno,bend-shinoda}. 

\begin{center}
\includegraphics[width=0.48\textwidth, angle=0]{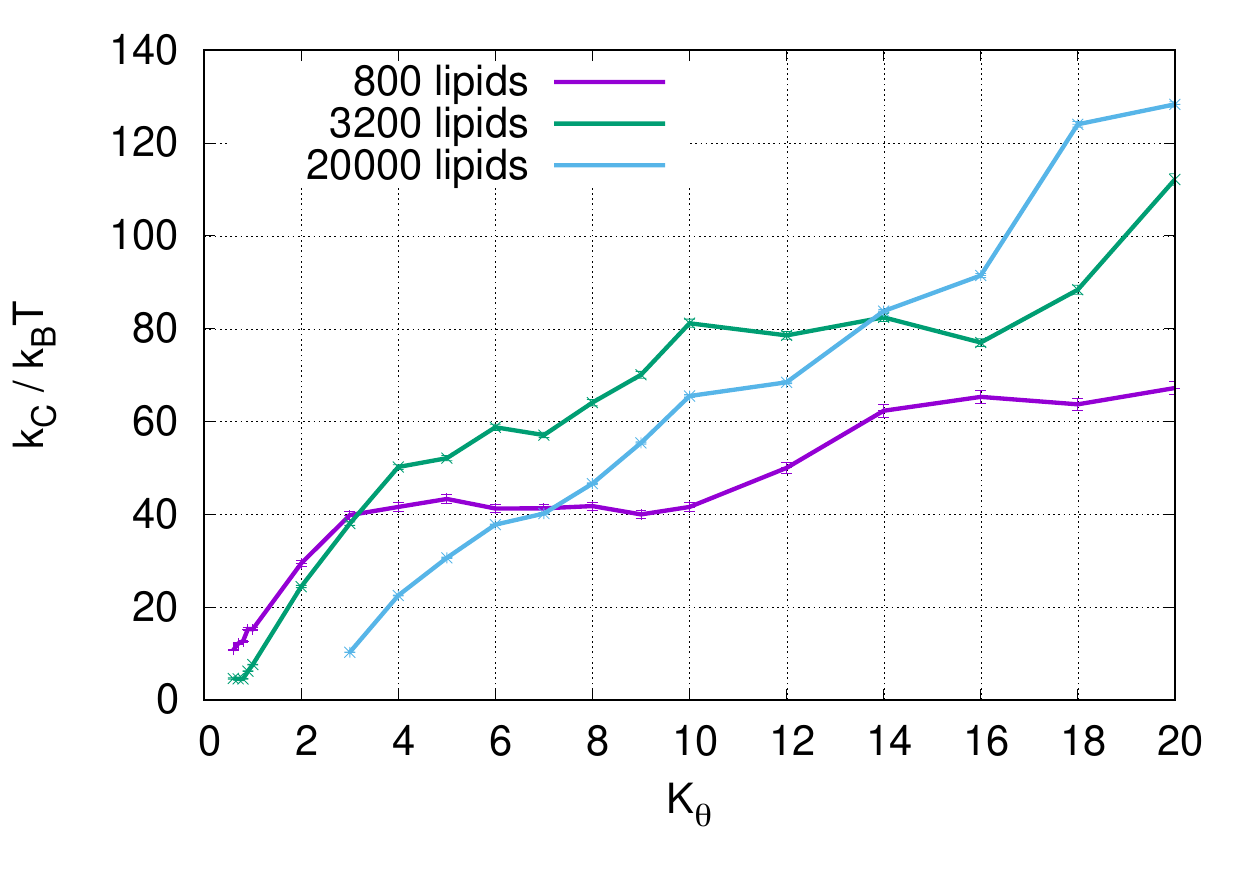}
\captionof{figure}{Bending modulus, $k_C$, plotted for the 4-bead lipid model as function of the angle potential parameter, $K_\theta$, for three different system sizes (800 lipids, 3200 lipids, and 20,000 lipids). Plottet errorbars indicate the statistical error in the least squares fit. Simulation time was $50\times 10^6$ MD steps.}
\label{fig:4site-bend-size}
\end{center}

\noindent {\bf The quasi-monolayer model}

Up until this point we have been concerned with describing properties of assemblies, and especially bilayers, consisting of linear chain molecular lipids with 2-5 bead per molecule (four left-most of Fig. \ref{fig:models}). Now let us have a look at a bilayer membrane representation that is not resolved at the level of individual lipids, but where a patch of membrane consists of a 3-bead segment. This quasi-monolayer is the most simple particle-based membrane representation we propose. Perhaps surprisingly, the quasi-monolayer model again exhibits fully-competent assembly behavior consistent with all other models (Fig. \ref{fig:quasi-assembly}) and tunable bilayer bending modulus all the way down to the order of the thermal energy, $k_BT$. 

\begin{center}
\includegraphics[width=0.48\textwidth, angle=0]{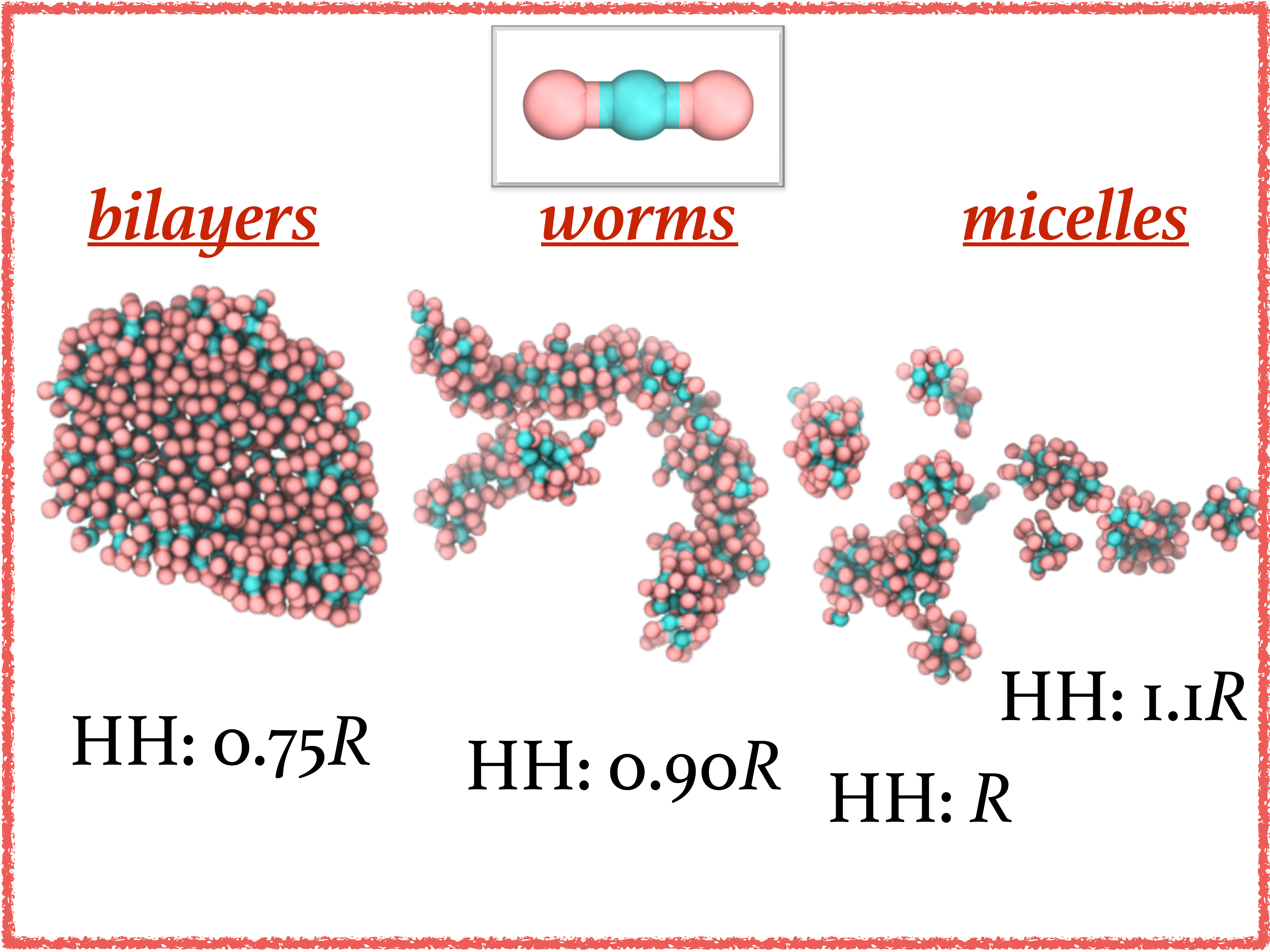}
\captionof{figure}{Self-assembled nano-structures of the quasi-monolayer model are conserved to those of bilayer models (see Fig. \ref{fig:assembly}). 225 lipids were simulated for $0.5\times 10^6$ MD steps. Movies corresponding to the snapshots are available in the supplemental material online \cite{supplstuff}.}
\label{fig:quasi-assembly}
\end{center}


In addition, we have performed \textit{in silico} pulling experiments to probe the flexibility of our model membranes and their ability to form long, thin "necks." Thin membrane "necks" are important in many aspects of biology e.g. endocytotic trafficking and viral budding. These experiments are akin to experimental assays where vesicles or cells are challenged by pulling using optical tweezers, nano-beads, or the micropipette aspiration technique. Our pulling method consists of a vesicle of diameter $D$ that is being pulled apart by two small, all-repulsive guiding potentials of diameter $d$. The guiding potentials are initialized within the vesicle and drawn out radially with a given pull rate, $|\vec{v}|$ (Fig. \ref{fig:pulling}a), which we set to $5 \text{ms}^{-1}$ or $10 \text{ms}^{-1}$ . We use our quasi-monolayer model with angle potential parameter of $K_\theta=\{2,20\} k_BT.\text{rad}^{-2}$ for this demonstration (Fig. \ref{fig:pulling}b). The simulations reveal that the membranes are extremely flexible and can form very thin "necks" of diameters less than $\sim 10$nm before membrane rupture, even at relatively fast pulling rates (Fig. \ref{fig:pulling}c). 

\begin{center}
\includegraphics[width=0.48\textwidth, angle=0]{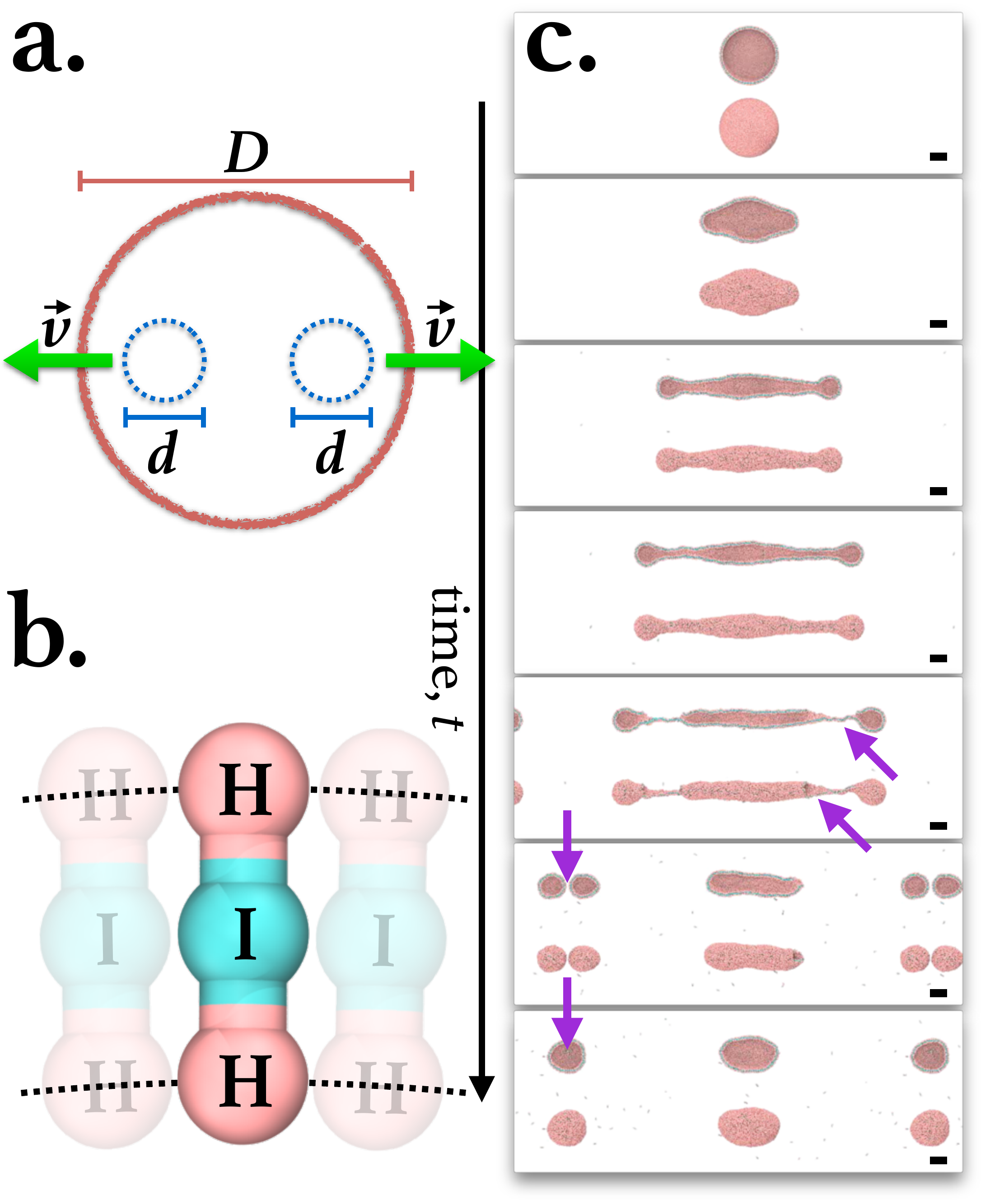}
\captionof{figure}{Pulling experiment setup. (a) Geometry description and indication of moving guiding potentials. (b) Drawing of the quasi-monolayer lipid model membrane is used in the pulling simulations. (c) Snapshots for the case where $D=40\text{nm}$, $d=10\text{nm}$, $|\vec{v}|=10\text{m s}^{-1}$, and $K_\theta=20 k_BT.\text{rad}^{-2}$. Top and bottom image in each snapshot is a rendering of the cross-section and surface views, respectively. The arrows indicate the location of membrane fission and fusion events. Scale bar indicates $10$nm. Movies corresponding to pulling speeds of the $|\vec{v}|=\{5,10\}\text{m s}^{-1}$ with membranes of lipids with $K_\theta=\{2,20\} k_BT.\text{rad}^{-2}$ are available in the supplemental material online \cite{supplstuff}. }
\label{fig:pulling}
\end{center}

\noindent {\bf Summary on models and how to choose the right one}

Bilayer membranes formed from the 5- and 4-bead models can have their bending properties effectively be tuned by the angle potential parameters of the lipids without changing the non-bonded interactions. In the fluid regime, we can achieve bending moduli ranging from less than $10k_BT$ to in excess of $100k_BT$. The membranes formed by 5- and 4-bead lipids are intuitively appealing, perhaps especially for inclusions such as transmembrane proteins because they can exhibit a (qualitatively/semi-quantitavely) correct lateral pressure profile \cite{bpb}. It is furthermore easy to construct membranes of mixed lipids by preserving the pair interactions for identical chemical groups. 
 
The 3-bead model is more efficient than both the 5- and the 4-bead models, but the bending modulus is best tuned through modifying the relative bead sizes and the modifying the non-bonded interaction parameters, which seems less chemically intuitive (at least to us). The same obviously goes for the 2-bead model, which has no angle potential at all. However, extending the basin of attraction of the non-bonded interaction potential is indeed a practical way of increasing the range of stability of fluid bilayer membranes and modulating the bending rigidity as demonstrated by Cooke \& Deserno \cite{cd}. 

The quasi-monolayer model is designed to eliminate the need for lipid flipflop relaxation to balance inter-monolayer pressures. In our experience, it works well with phenomenological models of proteins (e.g., elastic network models \cite{enm1,enm2,enm3,enm4}) that adsorb to the membrane surface (e.g., scaffolding proteins \cite{pak, jarin}) and, in certain circumstances, even models of transmembrane proteins could be successfully introduced \cite{enm5}. The quasi-monolayer model is preferred when it is desirable to have a self-assembling membrane capable of significant remodeling, but the molecular details of the membrane interior are not important. 

All models can in principle be customized by applying empirical or systematic corrections to the generic interaction potentials utilized. Popular methods for administering the latter include iterative Boltzmann inversion \cite{ibi}, inverse Monte Carlo/Newton inversion \cite{imc,newtoninv}, molecular renormalization group coarse-graining \cite{molrenorm1,molrenorm2}, force matching \cite{fm1,fm2,fm3}, and relative entropy minization \cite{rem}. However, all of them can benefit from initial-guess potentials that gives rise to self-assembly-competent models; a virtue offered at different resolutions by the models we describe here. Example properties that could be parametrized include lipid phase behavior, compositional heterogeneity in multicomponent membranes, and protein-lipid interactions with correct local structural correlations. Future work will explore this possibility in more detail.

\section{\label{sec:conclusion} IV. CONCLUSION}
To conclude, lipid models of intermediary resolution in-between the fully atomistic and continuum regimes (sometimes called mesoscopic or coarse-grained) provide a convenient description that can be used to answer biophysical questions about biological membranes and their processes.  We propose a framework of phenomenological lipid models in implicit solvent in the same vein as some earlier models but with tangible improvements, in particular computational efficiency and model tunability over vast scales and even resolutions. The efficiency of our model comes from the adaptation of soft, short-ranged potentials and results in the increased numerical stability and boosted dynamics. We demonstrate assembly competency, topological changes, membrane bending properties, and membrane phase behavior as a result of tuning the lipid flexibility for bilayer models with 2-5 beads per molecule as well as a highly coarse-grained quasi-monolayer model. Models of this kind are complementary to tried-and-true computational fluid dynamics (CFD) and atomistic molecular dynamics (MD) simulations, and as such can offer new insights into experimental observations and theoretical deliberations.

\section{\label{sec:appendix} APPENDIX}

\noindent {\bf A. Rationale}

The basic functional form for the pair potential use have chosen has three intuitive ingredients,
\begin{enumerate}
\item Bead size, $R$. 
\item Lipid cohesion (Fig. \ref{fig:rationale}). Proportional of the free-energy change, $\sim\Delta F$, of lipid transfer from the monomer fraction, which represents the solution, into the lipid assembly or aggregate. 
\item Lipid rigidity (\textit{via} the angle and bond potentials).
\end{enumerate}

The basic modular platform consists of
\begin{itemize}
\item "Head" beads: These represent the vicinity around/above e.g. the phospholipid head groups including tightly bound waters. It is softer than other beads, which incorporates the effects of local solvation (in an effective or \textit{ad hoc} sense. For a systematic approach to solvation see e.g. ref. \cite{virtsitelip}). The pair interactions are purely repulsive (toward other head beads; no interactions with other beads) and the effective H-H $R$ size controls the packing ratio of the lipid (i.e., how "conical" it is).
\item "Interface" beads: These are placed immediately below the head beads and represent the interface between solvent and the hydrophobic membrane interior. Interactions are chosen to have a finite, soft repulsion at small distances and short-ranged attractive basin averaged out on all interface+tail beads \footnote{Equal distribution of the lipid cohesion onto interface+tail beads is not required but chosen for simplicity. A larger proportion of the cohesion can be attributed to the interfacial beads e.g. if one desires to tune the lateral pressure profile of the membrane bilayer as done in the model by Brannigan et al. \cite{bpb,bb}.}.
\item "Tail" beads: These represent the hydrocarbon moieties of the membrane interior. Interactions are the same as interface type beads (see above).
\end{itemize}

\begin{center}
\includegraphics[width=0.48\textwidth, angle=0]{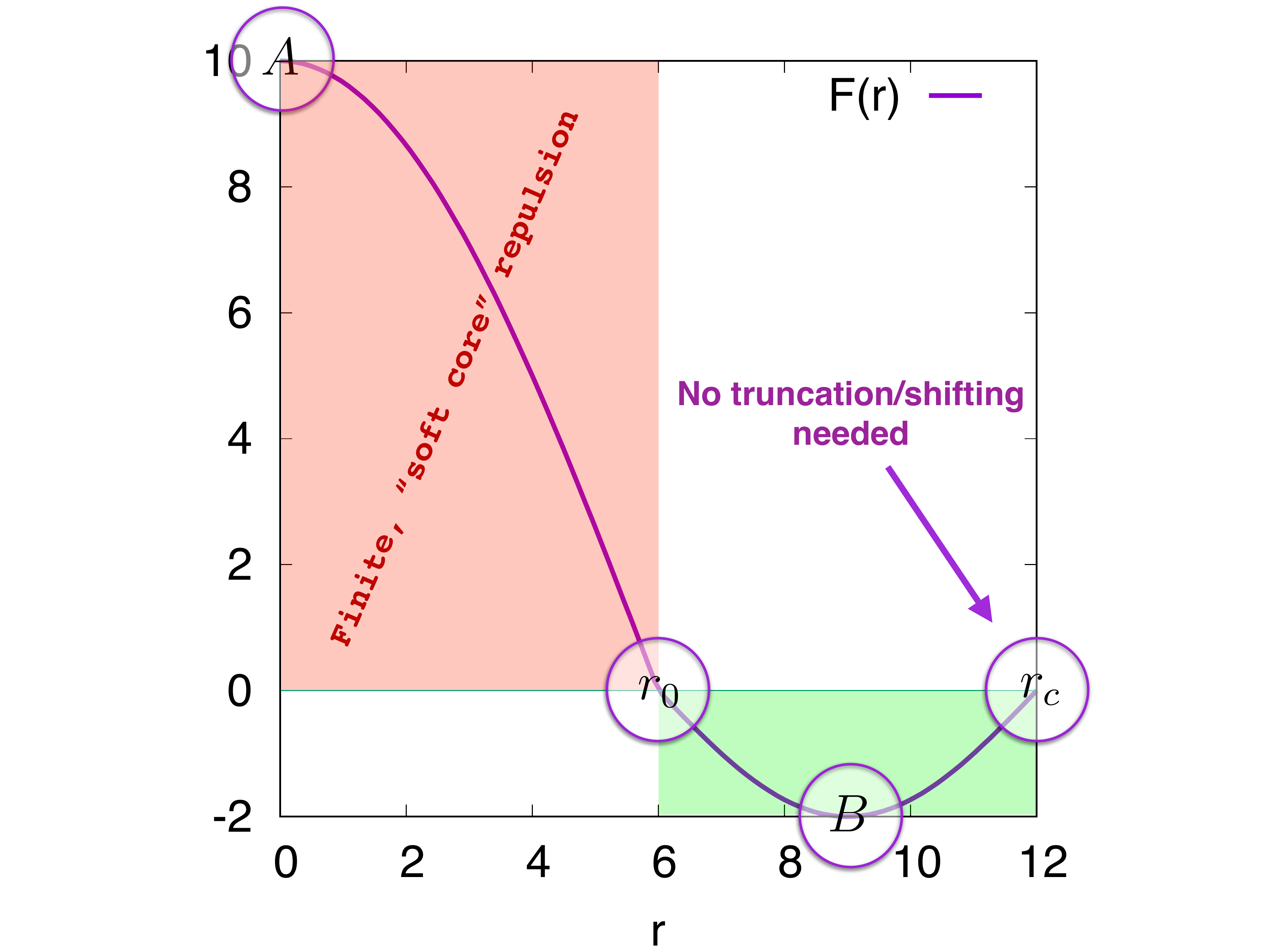}
\captionof{figure}{Example pairwise interaction force for the values $A=10$, $B=2$, $r_0=6$, and $r_c=12$ with visual emphasis on the interpretation of each parameter constant. The red-shaded region is the finite, soft-core repulsive regime and the green-shaded region is the attractive basin. The force vanishes smoothly as $r\rightarrow r_c^-$, so no truncation/shifting is needed.}
\label{fig:rationale}
\end{center}

\noindent {\bf B. Packaging and equilibration of bilayer membrane vesicles}

Several approaches are possible when packing lipids into a vesicle geometry. Typically, one would use an algorithm to generate (roughtly) equidistant points on the appropriate spheres (e.g., \cite{lloyd,cooksphere,marsagliasphere,muellersphere,rusinsphere,saffkuijlaars,spheredeserno, spheregonzalez}). Notice, however, that special care must be taken as there are several design choices available of varying practicality. Ideally, the packaged vesicle should not require flipflop relaxation because this can give rise to stability issues in our experience. To illustrate this, compare the resulting vesicles packed with constant volume per lipid (VPL) \textit{vs.} constant area per lipid (APL) with the spiral equidistant algorithm (Fig. \ref{fig:VPL}). The former yields a much more stable result, especially for small vesicles.

\begin{center}
\includegraphics[width=0.48\textwidth, angle=0]{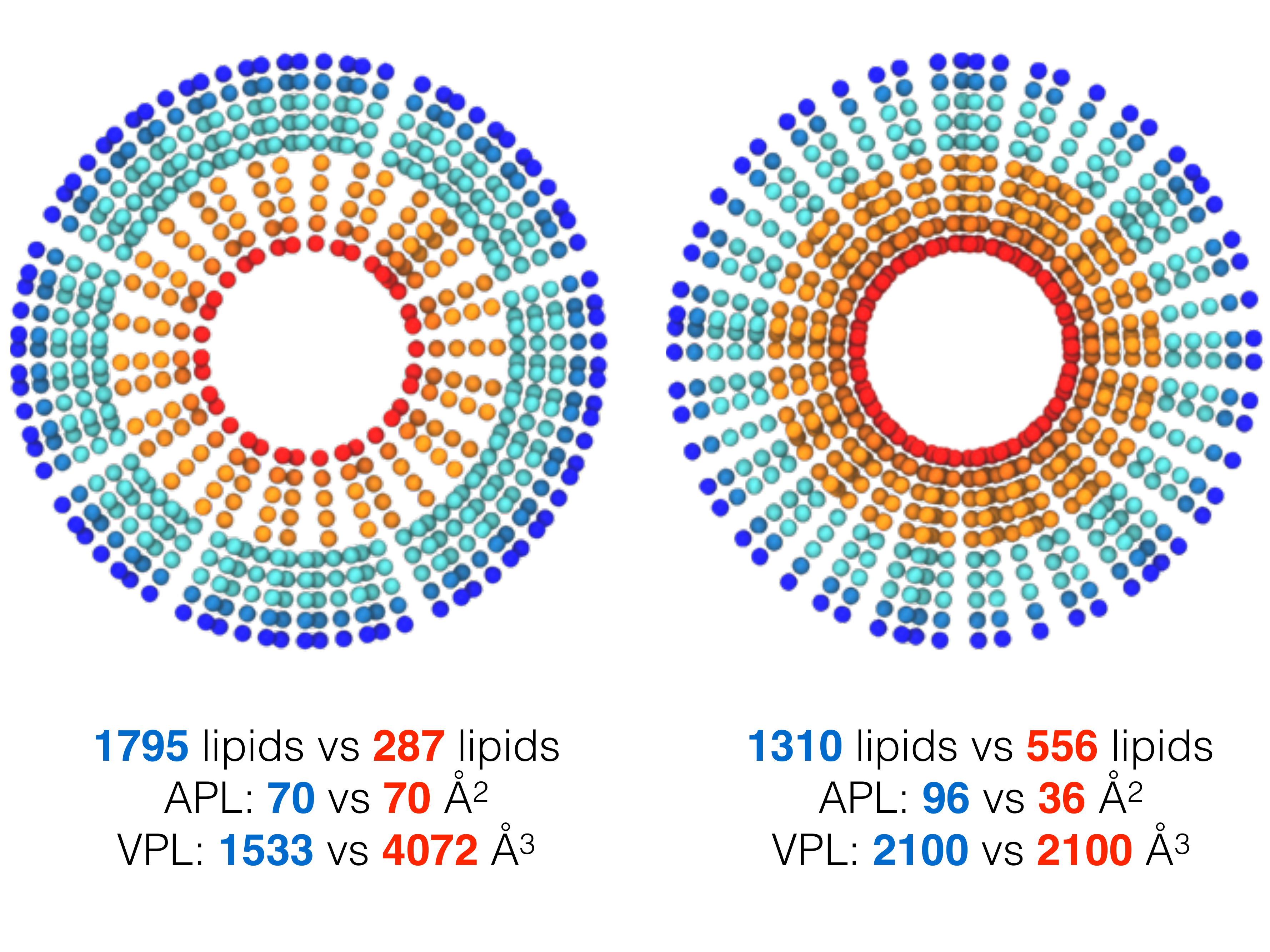}
\captionof{figure}{Cross-sectional slice of two vesicles that were packed with equal areas per lipid (\textit{left}) vs. equal volumes per lipid (\textit{right}) in the outer and inner leaflets. Visual inspection of the two packing strategies is illustrative, albeit somewhat misleading (e.g., there is no overlap between head group beads in the inner leaflet of the right figure because the closely positioned lipids are staggered in the out-of-plane dimension). }
\label{fig:VPL}
\end{center}




\section{\label{sec:acknowledgements} ACKNOWLEDGEMENTS}

We thank Gregory A. Voth and the members of his lab for stimulating scientific discussions while being at The University of Chicago, where the work presented in this article begun. The last-named author gratefully acknowledges financial support from the Carlsberg Foundation in the form of a postdoctoral fellowship (grants CF15-0552, CF16-0639, and CF17-0783) that in part financed said stay. 

%

\bibliography{mybib}{}

\begin{thebibliography}{10}

\bibitem{radha}
R.~Bradley and R.~Radhakrishnan, ``Coarse-grained models for protein-cell
  membrane interactions,'' {\em Polymers}, vol.~5, pp.~890--936, 2013.

\bibitem{farago}
O.~Farago, ``Water-free computer model for fluid bilayer membranes,'' {\em
  Journal of Chemical Physics}, vol.~119, pp.~596--605, 2003.

\bibitem{bpb}
G.~Brannigan, P.~F. Philips, and F.~L.~H. Brown, ``Flexible lipid bilayers in
  implicit solvent,'' {\em Physical Review E}, vol.~72, p.~011915, 2005.

\bibitem{ckd}
I.~R. Cooke, K.~Kremer, and M.~Deserno, ``Tunable generic model for fluid
  bilayer membranes,'' {\em Physical Review E}, vol.~72, p.~011506, 2005.

\bibitem{dpd1}
P.~Español, M.~Serrano, and I.~Zuñiga, ``Coarse-graining of a fluid and its
  relation with dissipative particle dynamics and smoothed particle dynamics,''
  {\em International Journal of Modern Physics C}, vol.~08, pp.~899--908, 1997.

\bibitem{dpd2}
E.~G. Flekkøy and P.~V. Coveney, ``From molecular dynamics to dissipative
  particle dynamics,'' {\em Physical Review Letters}, vol.~83, p.~1775, 1999.

\bibitem{dpd3}
P.~Español and P.~B. Warren, ``Perspective: Dissipative particle dynamics,''
  {\em Journal of Chemical Physics}, vol.~146, p.~150901, 2017.

\bibitem{dpd4}
Y.~Han, J.~F. Dama, and G.~A. Voth, ``Mesoscopic coarse-grained representations
  of fluids rigorously derived from atomistic models,'' {\em Journal of
  Chemical Physics}, vol.~149, p.~044104, 2018.

\bibitem{sdk}
W.~Shinoda, R.~DeVane, and M.~L. Klein, ``Multi-property fitting and
  parameterization of a coarse grained model for aqueous surfactants,'' {\em
  Molecular Simulation}, vol.~33, pp.~27--36, 2007.

\bibitem{mueller}
M.~H\"omberg and M.~M\"uller, ``Main phase transition in lipid bilayers: Phase
  coexistence and line tension in a soft, solvent-free, coarse-grained model,''
  {\em Journal of Chemical Physics}, vol.~132, p.~155104, 2010.

\bibitem{sunilkumar}
J.~D. Revalee, M.~Laradji, and P.~B.~S. Kumar, ``Implicit-solvent mesoscale
  model based in soft-core potentials for self-assembled lipid membranes,''
  {\em Journal of Chemical Physics}, vol.~128, p.~035102, 2008.

\bibitem{dpd}
P.~J. Hoogerbrugge and J.~M. V.~A. Koelman, ``Simulating microscopic
  hydrodynamic phenomena with dissipative particle dynamics,'' {\em Europhysics
  Letters}, vol.~19, pp.~155--160, 1992.

\bibitem{smit}
F.~de~Meyer and B.~Smit, ``Effect of cholesterol on the structure of a
  phospholipid bilayer,'' {\em Proceedings of the National Academy of Sciences
  of the United States of America}, vol.~106, pp.~3654--3658, 2009.

\bibitem{guo}
T.~Lu and H.~Guo, ``Phase behavior of lipid bilayers: A dissipative particle
  dynamics simulation study,'' {\em Advanced Theory and Simulations},
  p.~1800013, 2018.

\bibitem{Note1}
Parameters can obviously be chosen in order to make the potential $\protect
  \mathscr {C}^2$-continuous. While this is an unnecessarily strict restriction
  for practical applications, we do advice that care is taken if the
  discontinuity in the derivative of the force is.

\bibitem{mtk}
G.~J. Martyna, D.~J. Tobias, and K.~L. Klein, ``Constant pressure molecular
  dynamics algorithms,'' {\em Journal of Chemical Physics}, vol.~101, p.~4177,
  1994.

\bibitem{lenz}
O.~Lenz and F.~Schmid, ``A simple computer model for liquid lipid bilayers,''
  {\em Journal of Molecular Liquids}, vol.~117, pp.~147--152, 2005.

\bibitem{atzberger}
Y.~Wang, J.~K. Sigurdsson, E.~Brandt, and P.~J. Atzberger, ``Dynamic
  implicit-solvent coarse-grained models of lipid bilayer membranes:
  Fluctuating hydrodynamics thermostat,'' {\em Physical Review E}, vol.~88,
  p.~023301, 2013.

\bibitem{lammps}
S.~Plimpton, ``Fast parallel algorithms for short-range molecular dynamics,''
  {\em Journal of Computational Physics}, vol.~117, pp.~1--19, 1995.

\bibitem{supplstuff}
J.~M.~A. Grime and J.~J. Madsen, ``The grime coarse-grained lipid model,''
  2019.
\newblock [Online at \url{https://doi.org/10.5281/zenodo.3479543} 10-Oct-2019].

\bibitem{canham}
P.~B. Canham, ``The minimum energy of bending as a possible explanation of the
  biconcave shape of the human red blood cell,'' {\em Journal of Theoretical
  Biology}, vol.~26, pp.~61--81, 1970.

\bibitem{helfrich}
W.~Helfrich, ``Elastic properties of lipid bilayers: Theory and possible
  experiments,'' {\em Zeitschrift f\"ur Naturforschung}, vol.~28, pp.~693--703,
  1973.

\bibitem{spectra}
E.~G. Brandt, A.~R. Braun, J.~N. Sachs, J.~F. Nagle, and O.~Edholm,
  ``Interpretation of fluctuation spectra in lipid bilayer simulations,'' {\em
  Biophysical Journal}, vol.~100, pp.~2104--2111, 2011.

\bibitem{israel}
J.~Israelachvili, D.~J. Mitchell, and B.~W. Ninham, ``Theory of self-assembly
  of hydrocarbon amphiphiles into micelles and bilayers,'' {\em Journal of the
  Chemical Society, Faraday Transactions}, vol.~72, pp.~1525--1568, 1976.

\bibitem{nielsen}
W.~Shinoda, T.~Nakamura, and S.~O. Nielsen, ``Free energy analysis of
  vesicle-to-bicelle transformation,'' {\em Soft Matter}, vol.~7,
  pp.~9012--9020, 2011.

\bibitem{bend-rigatoni}
V.~A. Harmandaris and M.~Deserno, ``A novel method for measuring the bending
  rigidity of model lipid membranes by simulating tethers,'' {\em Journal of
  Chemical Physics}, vol.~125, p.~204905, 2006.

\bibitem{bend-deserno}
M.~Hu, P.~Diggins, and M.~Deserno, ``Determining the bending modulus of a lipid
  membrane by simulating buckling,'' {\em Journal of Chemical Physics},
  vol.~138, p.~214110, 2013.

\bibitem{bend-shinoda}
S.~Kawamoto, T.~Nakamura, S.~O. Nielsen, and W.~Shinoda, ``A guiding potential
  method for evaluating the bending rigidity of tensionless lipid membranes
  from molecular simulation,'' {\em Journal of Chemical Physics}, vol.~139,
  p.~034108, 2013.

\bibitem{cd}
I.~R. Cooke and M.~Deserno, ``Solvent-free model for self-assembling fluid
  bilayer membranes: Stabilization of the fluid phase based on broad attractive
  tail potentials,'' {\em Journal of Chemical Physics}, vol.~123, p.~224710,
  2005.

\bibitem{enm1}
M.~M. Tirion, ``Large amplitude elastic motions in proteins from a
  single-parameter, atomic analysis,'' {\em Physical Review Letters}, vol.~77,
  pp.~1905--1908, 1996.

\bibitem{enm2}
T.~Haliloglu, I.~Bahar, and B.~Erman, ``Gaussian dynamics of folded proteins,''
  {\em Physical Review Letters}, vol.~79, pp.~3090--3093, 1997.

\bibitem{enm3}
J.~J. Madsen, A.~V. Sinitskiy, J.~Li, and G.~A. Voth, ``Highly coarse-grained
  representations of transmembrane proteins,'' {\em Journal of Chemical Theory
  and Computations}, vol.~13, pp.~935--944, 2017.

\bibitem{enm4}
A.~V. Sinitskiy and G.~A. Voth, ``Coarse-graining of proteins based on elastic
  network models,'' {\em Chemical Physics}, vol.~422, pp.~165--174, 2013.

\bibitem{pak}
A.~J. Pak, J.~M.~A. Grime, P.~Sengupta, A.~K. Chen, A.~E.~P. Durumeric,
  A.~Srivastava, M.~Yaeger, J.~A.~G. Briggs, J.~Lippincott-Schwartz, and G.~A.
  Voth, ``Immature hiv-1 lattice assembly dynamics are regulated by scaffolding
  from nucleic acid and the plasma membrane,'' {\em Proceedings of the National
  Academy of Sciences of the United States of America}, vol.~114,
  pp.~E10056--E10065, 2017.

\bibitem{jarin}
Z.~Jarin, F.-C. Tsai, A.~Davtyan, A.~J. Pak, P.~Bassereau, and G.~A. Voth,
  ``Unusual organization of i-bar proteins on tubular and vesicular
  membranes,'' {\em Biophysical Journal}, vol.~117, pp.~553--562, 2019.

\bibitem{enm5}
J.~J. Madsen, J.~M.~A. Grime, J.~S. Rossman, and G.~A. Voth, ``Entropic forces
  drive clustering and spatial localization of influenza a m2 during viral
  budding,'' {\em Proceedings of the National Academy of Sciences of the United
  States of America}, vol.~115, pp.~E8595--E8603, 2018.

\bibitem{ibi}
D.~Reith, M.~P\"utz, and F.~M\"uller-Plathe, ``Deriving effective mesoscale
  potentials from atomistic simulations,'' {\em Journal of Computational
  Physics}, vol.~24, pp.~1624--1636, 2003.

\bibitem{imc}
A.~P. Lyubartsev and A.~Laaksonen, ``Calculation of effective interaction
  potentials from radial distribution functions: A reverse monte carlo
  approach,'' {\em Physical Review E}, vol.~52, pp.~3730--3737, 1995.

\bibitem{newtoninv}
A.~Na\^om\'e, A.~Laaksonen, and D.~P. Vercauteren, ``A solvent-mediated
  coarse-grained model of dna derived with the systematic newton inversion
  method,'' {\em Journal of Chemical Theory and Computation}, vol.~10,
  pp.~3541--3549, 2014.

\bibitem{molrenorm1}
A.~Savelyev and G.~A. Papoian, ``Molecular renormalization group
  coarse-graining of polymer chains: application to double-stranded dna,'' {\em
  Biophysical Journal}, vol.~96, pp.~4044--4052, 2009.

\bibitem{molrenorm2}
A.~Savelyev and G.~A. Papoian, ``Molecular renormalization group
  coarse-graining of electrolyte solutions: application to aqueous nacl and
  kcl,'' {\em Journal of Physical Chemistry B}, vol.~113, pp.~7785--7793, 2009.

\bibitem{fm1}
F.~Ercolessi and J.~B. Adams, ``Interatomic potentials from first-principles
  calculations: The force-matching method,'' {\em Europhysics Letters (EPL)},
  vol.~26, pp.~583--588, 1994.

\bibitem{fm2}
S.~Izvekov and G.~A. Voth, ``A multiscale coarse-graining method for
  biomolecular systems,'' {\em Journal of Physical Chemistry B}, vol.~109,
  pp.~2469--2473, 2005.

\bibitem{fm3}
W.~G. Noid, J.~W. Chu, G.~S. Ayton, V.~Krishna, S.~Izvekov, G.~A. Voth, A.~Das,
  and H.~C. Andersen, ``The multiscale coarse-graining method. i. a rigorous
  bridge between atomistic and coarse-grained models,'' {\em Journal of
  Chemical Physics}, vol.~128, p.~244114, 2008.

\bibitem{rem}
M.~S. Shell, ``The relative entropy is fundamental to multiscale and inverse
  thermodynamic problems,'' {\em Journal of Chemical Physics}, vol.~129,
  p.~144108, 2008.

\bibitem{virtsitelip}
A.~J. Pak, T.~Dannenhofer-Lafage, J.~J. Madsen, and G.~A. Voth, ``Systematic
  coarse-grained lipid force-fields with semi-explicit solvation via virtual
  sites,'' {\em Journal of Chemical Theory and Computation}, vol.~15,
  pp.~2087--2100, 2019.

\bibitem{Note2}
Equal distribution of the lipid cohesion onto interface+tail beads is not
  required but chosen for simplicity. A larger proportion of the cohesion can
  be attributed to the interfacial beads e.g. if one desires to tune the
  lateral pressure profile of the membrane bilayer as done in the model by
  Brannigan et al. \cite {bpb,bb}.

\bibitem{lloyd}
S.~P. Lloyd, ``Least squares quantization in pcm,'' {\em IEEE Transactions on
  Information Theory}, vol.~28, pp.~129--137, 1982.

\bibitem{cooksphere}
J.~M. Cook, ``Technical notes and short papers: Rational formulae for the
  production of a spherically symmetric probability distribution,'' {\em
  Mathematical Tables and Other Aids to Computation}, vol.~11, pp.~81--82,
  1957.

\bibitem{marsagliasphere}
G.~Marsaglia, ``Choosing a point from the surface of a sphere,'' {\em The
  Annals of Mathematical Statistics}, vol.~43, pp.~645--646, 1972.

\bibitem{muellersphere}
M.~E. Muller, ``A note on a method for generating points uniformly on
  n-dimensional spheres,'' {\em Communications of the Association for Computing
  Machinery}, vol.~2, pp.~19--20, 1959.

\bibitem{rusinsphere}
D.~Rusin, ``N-dim spherical random number drawing,'' {\em The Mathematical
  Atlas}.
\newblock [Online at
  \url{http://www.math.niu.edu/~rusin/known-math/96/sph.rand}].

\bibitem{saffkuijlaars}
E.~B. Saff and A.~B.~J. Kuijlaars, ``Distributing many points on a sphere,''
  {\em The Mathematical Intelligencer}, vol.~19, pp.~5--11, 1997.

\bibitem{spheredeserno}
M.~Deserno, ``How to generate equidistributed points on the surface of a
  sphere,'' 2004.
\newblock [Online at
  \url{https://www.cmu.edu/biolphys/deserno/pdf/sphere_equi.pdf} 10-Oct-2019].

\bibitem{spheregonzalez}
A.~Gonz\'alez, ``Measurement of areas on a sphere using fibonacci and
  latitude-longitude lattices,'' {\em arXiv:0912.4540}, pp.~1--19, 2009.

\bibitem{bb}
G.~Brannigan and F.~L.~H. Brown, ``A consistent model for thermal fluctuatins
  and protein-induced deformations in lipid bilayers,'' {\em Biophysical
  Journal}, vol.~90, pp.~1501--1520, 2006.

\end{thebibliography}
\bibliographystyle{ieeetr}

\end{document}